\documentclass[%
reprint,
amsmath,amssymb,
aps,
nolongbibliography
]{revtex4-2}

\usepackage{graphicx}
\usepackage{dcolumn}
\usepackage{bm}
\usepackage[colorlinks=true, allcolors=blue]{hyperref}
\usepackage{xcolor}
\usepackage{soul}

\usepackage{physics}
\usepackage{amsmath}
\usepackage{amssymb}
\usepackage{mathrsfs}
\allowdisplaybreaks

\usepackage{pdfpages}
\usepackage{pgffor}

\makeatletter
\AtBeginDocument{\let\LS@rot\@undefined}
\makeatother

\begin{document}
	
	
	\title{Instantaneous tunneling of relativistic massive spin-0 particles}
	
	\author{Philip Caesar Flores}
	\email{pmflores2@up.edu.ph}
	\author{Eric A. Galapon}%
	\email{eagalapon@up.edu.ph}
	\affiliation{%
		Theoretical Physics Group, National Institute of Physics \\ 
		University of the Philippines Diliman, 1101 Quezon City, Philippines
	}%
	
	%
	%
	
	\date{\today}

	\begin{abstract}
		The tunneling time problem earlier studied in \href{https://link.aps.org/doi/10.1103/PhysRevLett.108.170402}{Phys. Rev. Lett. \textbf{108}, 170402 (2012)} using a non-relativistic time-of-arrival (TOA) operator predicted that tunneling time is instantaneous. This raises the question on whether instantaneous tunneling time is a consequence of using a non-relativistic theory. Here, we extend the analysis by proposing a formalism on the construction of relativistic TOA-operators for spin-0 particles in the presence of an interaction potential $V(q)$ via quantization. We then construct the corresponding barrier traversal time operator, and impose the condition that the barrier height $V_o$ is less than the rest mass energy of the particle. We show that only the above-barrier energy components of the incident wavepacket’s momentum distribution contribute to the barrier traversal time while the below-barrier components are transmitted instantaneously.
	\end{abstract}
	
	\maketitle
	
	
	The problem of how long a particle tunnels through the classically forbidden region of a potential barrier is known as the quantum tunneling time problem. However, there are various conflicting theories regarding the problem as time is not an observable in standard quantum mechanics (SQM),  as such, the problem may be ill-defined as SQM has no canonical formalism to answer questions regarding time durations \cite{LANDSMAN20151}. This has lead to several definitions of tunneling time using a parametric approach, e.g. Wigner phase time, B\"{u}ttiker-Landauer time, Larmor time, Pollak-Miller time, among many others \cite{RevModPhys.61.917,RevModPhys.66.217,PhysRev.98.145,buttiker1982traversal,buttiker1983larmor,landauer1994barrier,PhysRevA.36.4604,yamada2004unified,pollak1984new,hartman1962tunneling,PhysRevLett.108.170402,brouard1994systematic,jaworski1988time}. The relation between these various proposed tunneling times is still unclear but can be classified into two distinct categories \cite{PhysRevLett.127.133001}, i.e. ``arrival times'' and ``interaction times''. The difference is that ``arrival times'' are concerned with the appearance of the tunneled particle on the far side of the barrier while ``interaction times'' determine the duration that the particle has spent inside the barrier \cite{PhysRevLett.127.133001}. This has helped clear opposing results of recent experiments done by Refs. \cite{sainadhAttosecondAngularStreaking2019a} and \cite{ramosMeasurementTimeSpent2020} wherein the former demonstrates an instantaneous tunneling  which uses an ``arrival time'' while the latter demonstrates a non-zero tunneling time via an ``interaction time''.  
	
	Earlier, one of us has addressed the tunneling time problem using a time-of-arrival (TOA) operator for a square potential barrier. By doing so, tunneling time is treated as a dynamical observable which addresses any contentions on tunneling time being an ill-defined problem. It was shown that only the above-barrier energy components of the initial wavefunction contribute to the barrier traversal time while below-barrier components are transmitted instantaneously, implying that tunneling is instantaneous. This supports the results of Ref. \cite{sainadhAttosecondAngularStreaking2019a} and earlier experiments done in Refs. \cite{eckleAttosecondAngularStreaking2008,pfeifferAttoclockRevealsNatural2012} but later experiments \cite{Landsman:14,PhysRevLett.119.023201} involving multi-electron atoms support non-zero tunneling times. However, whether tunneling as an ``arrival time'' is instantaneous or not, the crux of the problem is that both results imply that the particle exhibits superluminal behavior below the barrier. This now raises the question on whether the superluminality is a consequence of using non-relativistic quantum mechanics, i.e., could there be a fundamental difference if one uses a relativistic theory? There have been several studies to extend the analysis of tunneling times (parametric approach of time) to the relativistic case using the Dirac equation \cite{de2007dirac,de2013study,PhysRevA.67.012110,krekora2001effects,PhysRevB.40.5387} and it was shown there that the superluminal behavior still persists.
	
	It is well-known that relativistic quantum mechanics is not a well-defined one-particle theory, and one might instead appeal to quantum field theory as proposed by Ref. \cite{martin2020quantum}. However, if we treat tunneling as a TOA problem, then the effects of spontaneous pair-creation and annihilation will render the concept of TOA meaningless, i.e., we are not sure if the particle that tunneled and arrived is the same particle we initially had. In this Letter, we address the superluminal behavior by extending the analysis of Ref. \cite{PhysRevLett.108.170402} using a relativistic one-particle TOA-operator for spin-0 particles. We choose a square potential barrier of length $L$ and impose the condition that the strength of the barrier $V_o$ is less than the rest mass energy to minimize the effects of spontaneous pair-creation and annihilation, in order to keep the treatment as a one-particle theory. We will show, within the constraints of this work, that the conclusion of Ref. \cite{PhysRevLett.108.170402} also holds in the relativistic case, i.e. tunneling is instantaneous at least in the context of ``arrival times''.       
	
	\begin{figure*}[t!]
		\centering
		\includegraphics[width=0.9\textwidth]{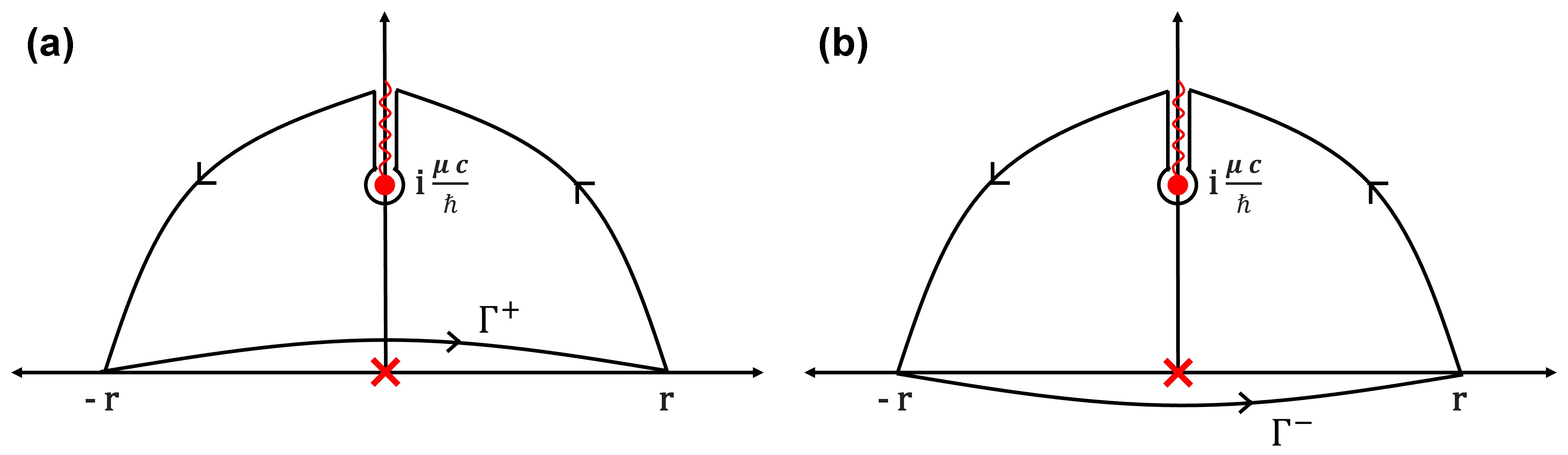}
		\caption{ Contours of integration for Eq. \eqref{eq:pkernel0SUPP} when $q-q' > 0$}
		\label{fig:contour}
	\end{figure*}
	
	Now, there is no existing theory on relativistic TOA-operators for an arbitrary potential $V(q)\neq0$ but there have been attempts to construct a relativistic free TOA-operator. The earliest was
	proposed by Razavi \cite{razavyQuantummechanicalConjugateHamiltonian1969a} which is a relativistic version of the Aharonov-Bohm TOA-operator \cite{PhysRev.122.1649}, and studied in detail by Ref. \cite{PhysRevA.105.062208}. We will thus first propose a formalism on the construction of relativistic TOA-operators via quantization using Ref. \cite{PhysRevA.105.062208,galapon2018quantizations,galapon2016cauchy,bender1988polynomials,tica2019finite}, and then construct the corresponding barrier traversal time operator for  the first time here.

	The relativistic TOA-operators are constructed by quantizing the ``classical'' relativististic TOA (CRTOA)
	\begin{equation}
		t_x(q,p) = -\text{sgn}p\int_x^q \dfrac{dq'}{c} \left(1-\dfrac{\mu^2c^4}{(H(q,p)-V(q'))^2}\right)^{-1/2}
		\label{eq:classreltoa}
	\end{equation} 
	which is obtained by inverting the Hamiltonian of special relativity $H(q,p) = \sqrt{p^2c^2+\mu^2c^4} + V(q)$, wherein, $x$ is the arrival point, $\mu$ is the rest mass, and $\text{sgn}(z)$ is the signum function. Now, in non-relativistic quantum mechanics, there is still no consensus on the quantization classical TOA for the interacting case $V(q)\neq0$ because it can be complex and/or multiple-valued. However, it has been argued that it is only meaningful to quantize the first TOA, and that the TOA of a quantum particle is always real-valued because it can tunnel through the classically forbidden region \cite{galapon2018quantizations}. We shall use use the same arguments to justify the quantization of Eq. \eqref{eq:classreltoa}.

	The classical expression Eq. \eqref{eq:classreltoa} is quantized in position representation using a modified Weyl-ordering rule \cite{galapon2018quantizations,bender1988polynomials,galapon2016cauchy,tica2019finite}. We first recast the classical expression Eq. \eqref{eq:classreltoa} into a form amenable to quantization by expanding around the free TOA 
	\begin{align}
		t_x(q,p)=& - \mu \sum_{j=0}^\infty \sum_{k=0}^j \binom{-\frac{1}{2}}{j} \binom{j}{k} \dfrac{(2\mu)^j \tilde{t}_{j,k}(q,p)}{(2\mu c^2)^{j-k}} 
		\label{eq:expand2quant} 
	\end{align}
	where, 
	\begin{align}
		\tilde{t}_{j,k}(q,p) = \int_x^q dq'  \sqrt{1 + \dfrac{p^2}{\mu^2 c^2}}^{k+1}  \dfrac{(V(q)-V(q'))^{2j-k}}{p^{2j+1}}.
		\label{eq:expand2quantfactor}
	\end{align} 
	The expansion Eq. \eqref{eq:expand2quant} is single and real-valued within its region of convergence in the phase space \cite{galapon2018quantizations}. Without loss of generality, we let the arrival point be the origin $x=0$, and assume that the potential $V(q)$ is analytic at the origin such that it admits the expansion $V(q)=\sum_{n=0}\nu_n q^n$. It follows that, 
	\begin{equation}
		\int_0^q dq' (V(q)-V(q'))^{u} = \sum_{n=1}^{\infty} a_n^{(u)} q^n.
		\label{eq:Vexpand},
	\end{equation}
	which turns Eq. \eqref{eq:expand2quantfactor} into
	\begin{align}
		\tilde{t}_{j,k}(q,p) = \sum_{n=1}^\infty a_n^{(2j-k)} \dfrac{q^n}{p^{2j+1}} \sqrt{1 + \dfrac{p^2}{\mu^2 c^2}}^{k+1}
		\label{eq:expand2quantfactora}
	\end{align}
	and is now amenable to quantization by promoting the position and momentum in $\tilde{t}_{j,k}(q,p)$ into operators $\mathsf{\hat{q}}$ and $\mathsf{\hat{p}}$. This is done by generalizing the Bender-Dunne basis operators \cite{bender1988polynomials} to any separable classical function $f(q,p)=g(q)^nh(p)^m$ such that 
	\begin{equation}
		\mathsf{\hat{f}_{\hat{q},\hat{p}}} = \dfrac{\sum_{k=0}^n \alpha_k^{(n)} \mathsf{\hat{g}_{\hat{q}}}^k\mathsf{\hat{h}_{\hat{p}}}^m\mathsf{\hat{g}_{\hat{q}}}^{n-k}}{\sum_{k=0}^n \alpha_k^{(n)}}. 
		\label{eq:BDbasisGen}
	\end{equation}
	Weyl-ordering is then imposed by choosing the coefficient $\alpha_k^{(n)}=n!/k!(n-k)!$. 
	
	In position representation, the TOA-operator becomes an integral operator
	\begin{equation}
		(\mathsf{\hat{T}}\psi)(q)=\int_{-\infty}^\infty dq' \mel{q}{\mathsf{\hat{T}}}{q'}\psi(q')
		\label{eq:timekernel}
	\end{equation}
	wherein the kernel is given as 
	\begin{align}
		\mel{q}{\mathsf{\hat{T}}}{q'} = - \mu & \sum_{j=0}^\infty \sum_{k=0}^j \binom{-\frac{1}{2}}{j} \binom{j}{k} \dfrac{(2\mu)^j}{(2\mu c^2)^{j-k}} \nonumber \\
		&\times \sum_{n=1}^\infty a_n^{(2j-k)}  \dfrac{(q+q')^n}{2^n} \nonumber \\
		&\times \mel{q}{\mathsf{\hat{p}}^{-2j-1} \sqrt{1 + \dfrac{\mathsf{\hat{p}}^2}{\mu^2 c^2}}^{k+1} }{q'}.
		\label{eq:timekernelexpandSUPP}
	\end{align}
	The momentum kernel is then evaluated by inserting the resolution of the identity $\mathsf{1}=\int_{-\infty}^{\infty} dp \ket{p}\bra{p}$ and using the plane wave expansion $\braket{q}{p}=e^{iqp/\hbar}/\sqrt{2\pi\hbar}$ which yields the expression 
	\begin{align}
		&\mel{q}{\mathsf{\hat{p}}^{-2j-1} \sqrt{1 + \dfrac{\mathsf{\hat{p}}^2}{\mu^2 c^2}}^{k+1} }{q'} \nonumber \\
		&=  \int_{-\infty}^{\infty} \frac{dp}{2\pi\hbar} \exp[\frac{i}{\hbar}(q-q')p] \frac{1}{p^{2j+1}} \left(  \sqrt{1 + \frac{p^2}{\mu^2 c^2}} \right)^{k+1}
		\label{eq:pkernel0SUPP}
	\end{align}
	The integral on the right hand side of Eq. \eqref{eq:pkernel0SUPP} diverges because of the pole of order $2j+1$ at $p=0$, but one of us has shown \cite{galapon2016cauchy} that we can assign a value to this divergent integral. This is done by interpreting the integral as a distribution and is evaluated by taking the average of the contour integrals $\int_{\Gamma^\pm} dz f(z) z^{-2j-1}$, where the path $\Gamma^+ (\Gamma^-)$ extends to $\pm\infty$ by going above (below) the pole at $z=0$. The resulting value coincides with the Cauchy principal value for $j=0$ and to the Hadamard finite part for non-negative integer values of $j$. Ref. \cite{PhysRevA.105.062208} has already evaluated the case when $j=k=0$, and we can use exactly the same contour, shown in Fig. \ref{fig:contour}, to generalize their method for non-negative integers $\{j,k\}$. 
	
	It can be seen that taking the average of the integrals along the paths $\Gamma^+$ and $\Gamma^-$in the limit as $r\rightarrow\infty$, only the contributions of the residue at $z=0$ and branch cut $(i\mu c , i \infty)$ will remain. The same method can be easily extended to the case $q-q'<0$ by closing the contour on the lower half of the complex plane and avoid the branch cut at $(-i\mu c , - i \infty)$. Performing the integration and substituting the resulting expression to Eq. \eqref{eq:timekernelexpandSUPP} yields
	\begin{equation}
		\mel{q}{\mathsf{\hat{T}}}{q'} = \dfrac{\mu}{i\hbar} T(q,q') \text{sgn}(q-q').
	\end{equation}
	
	\begin{widetext}	
		\noindent We refer to $T(q,q')$ as the time kernel factor (TKF) and has the form
		\begin{align}
			\tilde{T}(\eta,\zeta) =& \dfrac{1}{2} \int_0^{\eta}   ds \mathcal{F}(\eta,\zeta,s) + \dfrac{1}{\pi} \int_0^{\eta} ds \int_1^\infty dy  \exp[-\dfrac{\mu c}{\hbar}\abs{\zeta}y] \dfrac{\sqrt{y^2-1}}{y} \mathcal{G}(\eta,s,y), \label{eq:timekernelfunction} \\
			\mathcal{F}(\eta,\zeta,s) =& \int_0^\infty dy e^{-y} \oint_R \dfrac{dz}{2\pi i}  \dfrac{1}{z} \sqrt{1 + \dfrac{z^2}{\mu^2 c^2}}{_0}F_1 \left[ ; 1; \dfrac{\mu V(\eta,s)}{2\hbar^2}  \left( \zeta - i\hbar \dfrac{y}{z} \right)^2 \left( \sqrt{1 + \dfrac{z^2}{\mu^2 c^2}} + \dfrac{V(\eta,s)}{2\mu c^2}  \right) \right] \label{eq:tkf1} \\
			\mathcal{G}(\eta,s,y)	=& \dfrac{1}{2}
			\left\{ \left[ 1 - \left( \dfrac{1}{y}\dfrac{V(\eta,s)}{\mu c^2 }\right)^2 + i \dfrac{2\sqrt{y^2-1}}{y^2} \left( \dfrac{V(\eta,s)}{\mu c^2} \right) \right]^{-1/2} + \mathsf{g}_{i\rightarrow-i}\right\}
			\label{eq:tkf2}
		\end{align}
		where, ${_0}F_1(;a;z)$ is a specific hypergeometric function, and $V(\eta,s)=V(\eta)-V(s)$. Here, we performed a change of variable from $(q,q')$ to $(\eta,\zeta)$, where, $\eta=(q+q')/2$ and $\zeta=q-q'$ such that $T(q,q')=\tilde{T}(\eta,\zeta)$. The contour $R$ in Eq. \eqref{eq:tkf1} is a circle of radius $r<\mu c$ in the complex-plane that encloses the pole at $z=0$ which is equal to the residue at the origin. Meanwhile, $\mathsf{g}_{i\rightarrow-i}$ denotes changing $i$ to $-i$ in the first term of Eq. \eqref{eq:tkf2}. In the limit as $c\rightarrow\infty$, Eq. \eqref{eq:timekernelfunction} reduces to the known TKF for non-relativistic Weyl-quantized TOA-operators constructed in Ref. \cite{galapon2018quantizations}. 
	\end{widetext}
	
	Let us now consider a wavepacket $\psi(q)$ initially centered at $q=q_o$ with average momentum $p_o$ that is launched at $t=0$ towards a detector located at the arrival point $x=0$. The wavepacket is subjected to an interaction potential $V(q)$ located between $q_o$ and the arrival point $x$ such that the tail of $\psi(q)$ does not ``leak'' into V(q). The readers are directed to Ref. \cite{PhysRevLett.108.170402} for a full account of the measurement scheme implemented in the model. The average traversal time across the barrier is deduced from the difference of the average TOA in the presence and absence of the barrier which is assumed to be the expectation value of the corresponding operator, i.e., $\Delta\bar{\tau}=\bar{\tau}_F-\bar{\tau}_B=\mel{\psi}{\mathsf{\hat{T}_F}}{\psi}-\mel{\psi}{\mathsf{\hat{T}_B}}{\psi}$, wherein the subscripts $F$ and $B$ indicates the absence and presence of the barrier, respectively. 
	
	In the absence of the barrier, the TKF is obtained by substituting $V(q)=0$ into Eq. \eqref{eq:timekernelfunction}, which easily simplifies to $\tilde{T}_F(\eta,\zeta) = \frac{\eta}{2} \mathcal{T}_F(\zeta)$, where, 
	\begin{equation}
		\mathcal{T}_F(\zeta) = 1 + \dfrac{2}{\pi}\int_1^\infty dz\dfrac{\sqrt{z^2-1}}{z} \exp[- \dfrac{\mu c}{\hbar}\abs{\zeta}z].
	\end{equation}
	The operator corresponding to the TKF $\tilde{T}_F(\eta,\zeta)$ coincides with the Rigged Hilbert space extension of Razavi's relativistic free TOA-operator studied in Ref. \cite{PhysRevA.105.062208}, where it was shown that the associated physical quantities are consistent with special relativity. 
	
	The TOA-operator  Eq. \eqref{eq:timekernel} was derived under the assumption that the interaction potential is analytic, but since the TKF Eq. \eqref{eq:timekernelfunction} is in integral form, then Eqs. \eqref{eq:tkf1}-\eqref{eq:tkf2} can be applied to piecewise potentials such as the square barrier. We justify this by establishing that in the limit as $\hbar\rightarrow0$, the operator for the square potential barrier reduces to its corresponding CRTOA, as will be shown in the discussion to follow. Let us place the barrier of length $L=a-b$ to the left side of the arrival point $q=0$, i.e. $V(q)=V_o>0$ for $a<q<b<0$ and zero outside the interval $(a,b)$. The TKF may then be obtained by mapping the potential $V(q)$ into three non-overlapping regions in the $\eta$-coordinate such that the arrival point is at $\eta=0$ and $V(\eta)=V_o$ for $a<\eta<b<0$ and zero outside the interval $(a,b)$.
	
	We now obtain the barrier TKF for each region. In Region I, $\eta>b$, it is easy to see that $V(\eta)=0$ for the entire integration region which then turns Eq. \eqref{eq:timekernelfunction} into
	\begin{equation}
		\tilde{T}_B^{(I)}(\eta,\zeta)=  \frac{\eta}{2} \mathcal{T}_F(\zeta).
	\end{equation}
	In Region II, $a\leq\eta\leq b$, we have $V(\eta)=V_o$ and it is necessary to split the integral in Eq. \eqref{eq:timekernelfunction} into two parts as $V(\eta')=0$ for $b<\eta'<0$ and $V(\eta')=V_o$ for $\eta<\eta'<b$. This yields 
	\begin{align}
		\tilde{T}_B^{(II)}(\eta,\zeta)= \frac{\eta+b}{2}\mathcal{T}_F(\zeta)-\frac{b}{2}\mathcal{T}_B(V_o,\zeta).
	\end{align}
	wherein 
	\begin{align}
		\mathcal{T}_B(V_o,\zeta) =& \mathcal{F}_B(V_o,\zeta) \\
		&+ \dfrac{2}{\pi}\int_1^\infty dy \dfrac{\sqrt{y^2-1}}{y}\exp[-\frac{\mu c}{\hbar} \abs{\zeta}y] \mathcal{G}_B(V_o,y) \nonumber .
	\end{align}
	The functions $\mathcal{F}_B(V_o,\zeta)$ and $\mathcal{G}_B(V_o,z)$ are explicitly given as
	\begin{align}
		\mathcal{F}_B(V_o,\zeta) =& \int_0^\infty dy e^{-y} \oint_R \dfrac{dz}{2\pi i}  \dfrac{1}{z} \sqrt{1 + \dfrac{z^2}{\mu^2 c^2}} \nonumber \\
		&\times {_0}F_1 \left[ ; 1; \dfrac{\mu V_o}{2\hbar^2}  \left( \zeta - i\hbar \dfrac{y}{z} \right)^2 \mathcal{P}_B(V_o,z)\right] \label{eq:bkfF} \\
		\mathcal{G}_B(V_o,y) =& \dfrac{1}{2} \left( 1 - \left(\dfrac{1}{y}\dfrac{V_o}{\mu c^2}\right)^2 + 2i \dfrac{\sqrt{y^2-1}}{y^2} \left( \dfrac{V_o}{\mu c^2} \right)  \right)^{-1/2} \nonumber \\
		& + g_{i\rightarrow-i}.
		\label{eq:bkfG}
	\end{align}
	in which, $\mathcal{P}_B(V_o,z) = \sqrt{1 + z^2/(\mu c)^2} + V_o/(2\mu c^2)$. Last, in Region III, $\eta<a$, $V(\eta)=V_o$ and we must split the integral Eq. \eqref{eq:timekernelfunction} into three parts as $V(\eta')=V_o$ for $a\leq\eta'\leq b$ and zero at $\eta>b$ or $\eta<a$. This then yields 
	\begin{align}
		\tilde{T}_B^{(III)}(\eta,\zeta) = \frac{\eta+L}{2}\mathcal{T}_F(\zeta)-\frac{L}{2}\mathcal{T}_B(-V_o,\zeta).
	\end{align}
	
	We prove that the free and the barrier TOA-operator corresponding to the TKFs $\tilde{T}_F(\eta,\zeta)$ and $\tilde{T}_B(\eta,\zeta)$, respectively, are the quantization of the free and the interacting CRTOA. This is done by taking the inverse Weyl-Wigner transform $\tilde{t}(q_o,p_o) = (\mu/i\hbar) \int_{-\infty}^{\infty} d\zeta e^{-ip_o\zeta/\hbar} \tilde{T}(q_o,\zeta)\text{sgn}(\zeta)$, where, $q_o$ and $p_o$ are the initial position and momentum, respectively. For the free case, the first term of the the Weyl-Wigner transform of $\tilde{T}_F(q_o,\zeta)$ is evaluated by taking the inverse of the distributional Fourier transform $\int_{-\infty}^{\infty} dx x^{-1} e^{i\sigma x} = i \pi \text{sgn}\sigma$ \cite{gel1964ov}. Meanwhile, the order of integration for the second term are interchanged, and the inner integral is then evaluated as a Laplace transform. The resulting expression is further evaluated using the integral identity 
	\begin{equation*}
		\int_1^\infty  \dfrac{\sqrt{z^2-1}}{z} \dfrac{a^2}{a^2 +b^2z^2} dz = \dfrac{\pi}{2} \left(-1 + \sqrt{1+\dfrac{a^2}{b^2}}\right),
		\label{eq:integral}
	\end{equation*}
	for all real $a,b$ \cite{PhysRevA.105.062208} which can also be obtained using the calculus of residues. Combining the results yields
	\begin{equation}
		\tilde{t}_F=-\mu \dfrac{q_o}{p_o} \sqrt{1+ \dfrac{p_o^2}{\mu^2c^2}}
	\end{equation}
	which is the known classical relativistic free TOA.
	
	For the barrier case, it is easy to show that the inverse Weyl-Wigner transform of the kernel corresponding to Region I will be equal to the free case $\tilde{t}_B^{(I)} = \tilde{t}_F$. In Region II, the first term of the Weyl-Wigner transform is evaluated by expanding the hypergeometric function in $\mathcal{F}_B(V_o,\zeta)$ using its power series representation. The orders of summation and integration are then interchanged to perform a term-by-term integration. The resulting series converges as long the initial energy of the particle is above the barrier height. After some tedious algebra, the summation will yield a term that exactly cancels the contribution of $\mathcal{G}_B(V_o,\zeta)$. Thus, the classical limit is
	\begin{align}
		\tilde{t}_B^{(II)} 	=& - \dfrac{\mu (q_o + b)}{p_o} \sqrt{1 + \dfrac{p_o^2}{\mu^2 c^2}}  \nonumber \\
		& + \dfrac{b}{c} \sqrt{\dfrac{1+\dfrac{p_o^2}{\mu^2c^2}}{\left( \sqrt{1+\dfrac{p_o^2}{\mu^2c^2}} + \dfrac{V_o}{\mu c^2}\right)^2-1}}. 
	\end{align}
	The first term of $\tilde{t}_B^{(II)}$ is the classical relativistic free TOA from the edge of the barrier to the origin while the second term is the traversal time on top of the barrier. Repeating the same steps, the Weyl-Wigner transform in Region III yields
	\begin{align}
		\tilde{t}_B^{(III)} =& - \dfrac{\mu (q_o + L)}{p_o} \sqrt{1 + \dfrac{p_o^2}{\mu^2 c^2}} \nonumber \\
		& + \dfrac{L}{c} \sqrt{\dfrac{1+\dfrac{p_o^2}{\mu^2c^2}}{\left( \sqrt{1+\dfrac{p_o^2}{\mu^2c^2}} - \dfrac{V_o}{\mu c^2}\right)^2-1}}. 
	\end{align}
	The first term of $\tilde{t}_B^{(III)}$ is the traversal time across the interaction free region while the second term is the traversal time across the barrier region. The Weyl-Wigner transforms $\tilde{t}_B^{(II)}$ and $\tilde{t}_B^{(III)}$ also coincide with CRTOA obtained from directly integrating Eq. \eqref{eq:classreltoa}. Thus, the TOA-operator Eq. \eqref{eq:timekernel} is indeed a quantization of the CRTOA.
	
	Now, the expectation value of the TOA-operator $\mathsf{\hat{T}}$ for an incident wavepacket $\psi(q)$ is given as 
	\begin{equation}
		\bar{\tau} = \dfrac{\mu}{i\hbar} \int_{-\infty}^\infty \int_{-\infty}^\infty dqdq'  \psi^*(q)  T(q,q') \text{sgn}(q-q') \psi(q')
	\end{equation}
	We let the initial wavepacket $\psi(q)$ have a momentum expectation value $\langle \mathsf{\hat{p}} \rangle = \hbar k_o$, or equivalently, a group velocity $\nu_o=\hbar k_o/\mu$ such that the wavepacket has the form $\psi(q)=e^{ik_oq}\varphi(q)$, where $\mel{\varphi}{\mathsf{\hat{p}}}{\varphi}=0$. It will be convenient to perform a change of variable from $(q,q')$ to $(\eta,\zeta)$, and work with the complex-valued TOA $\bar{\tau}^*$. The imaginary component of which, corresponds to the physical quantity, i.e. $\bar{\tau}=\Im[\bar{\tau}^*]$ where 
	\begin{align}
		\bar{\tau}^* = -\dfrac{2\mu}{\hbar} \int_{-\infty}^\infty d\eta&  \int_0^\infty d\zeta e^{i k_o \zeta } \tilde{T}(\eta,\zeta) \nonumber \\ 
		&\times \varphi^*\left(\eta-\frac{\zeta}{2}\right)\varphi\left(\eta+\frac{\zeta}{2}\right)
	\end{align}
	In the succeeding expressions, we indicate complex-valued quantities with an asterisk $*$, wherein their imaginary component corresponds to the physical quantity.
	
	The measurable quantity for deducing the barrier traversal time is then related to the complex-valued TOA via $\Delta\bar{\tau}=\Im[\Delta\bar{\tau}^*] = \Im[\bar{\tau}_F^* - \bar{\tau}_B^*]$. Directly substituting the free and barrier TKFs on $\bar{\tau}^*$, and taking the difference thus yields $\Delta\bar{\tau}^* = (L/\nu_o)Q_c^*-(L/\nu_o)R_c^*$, where, 
	\begin{align}
		Q_c^* =& k_o \int_0^\infty d\zeta e^{i k_o \zeta } \mathcal{T}_F(\zeta) \Phi(\zeta) \label{eq:QcRaw} \\
		R_c^* =& k_o \int_0^\infty d\zeta e^{i k_o \zeta } \mathcal{T}_B(-V_0,\zeta) \Phi(\zeta) \label{eq:RCRaw},
	\end{align}
	in which, $\Phi(\zeta) = \int_{-\infty}^\infty d\eta \varphi^*(\eta - \zeta/2) \varphi(\eta+\zeta/2)$. Here, we imposed that $\varphi(q)$ is initially located at the left side of the barrier whose support does not extend in the barrier region so as to use the barrier TKF corresponding to Region III. Furthermore, we assume that $\varphi(q)$ is infinitely differentiable.
	
	To gain a physical understanding of $Q_c^*$ and $R_c^*$, we investigate their asymptotic expansions in the high energy limit $k_o\rightarrow\infty$. Now, the expectation value of the relativistic free TOA-operator corresponding to $\tilde{T}_F(\eta,\zeta)$ was recently calculated in Ref. \cite{PhysRevA.105.062208} and shown that the leading term is just the CRTOA. Thus, it easily follows 
	\begin{equation}
		Q_c \sim \sqrt{1+ \dfrac{p_o^2}{\mu^2c^2}}
	\end{equation}
	which is just the relativistic correction to the non-relativistic free TOA. Next, notice that $R_c^*$ is a Fourier integral with respect to the asymptotic parameter $k_o$ which can be evaluated through repeated integration by parts, and using the same steps outlined earlier. Performing these operations will yield 
	\begin{equation}
		R_c \sim \dfrac{p_o}{\mu c} \sqrt{\dfrac{E_p^2}{(E_p - V_o)^2 - \mu^2 c^4}}
		\label{eq:RphysMeaning}
	\end{equation}
	where $E_p = \sqrt{p^2c^2 + \mu^2 c^4}$. It can be seen that $R_c$ is just the ratio of the energy of the incident particle and its energy above the barrier which leads us to the interpretation that $R_c$ is just an effective index of refraction (IOR) of the barrier with respect to the initial wavepacket. The same interpretation was made in the non-relativistic case for the square potential barrier and well \cite{PhysRevLett.108.170402,PhysRevA.101.022103}.
	
	We now establish the expected traversal time across the barrier region. To do so, it will be convenient to introduce the inverse Fourier transform of the wavepacket $\varphi(q) = (2\pi)^{-1} \int_{-\infty}^\infty d\tilde{k} e^{i \tilde{k} q} \phi(\tilde{k})$ such that $\Phi(\zeta)=\int_{-\infty}^\infty d\tilde{k} e^{i\zeta \tilde{k}} |\phi(\tilde{k})|^2$. Then by performing a change of variable $\tilde{k}=k-k_o$, it is easy to see that $\phi(k-k_o)$ is the Fourier transform of the full incident wavefunction $\psi(q)=e^{i k_o q}\varphi(q)$, i.e. $\phi(k-k_o)=\tilde{\psi}(k)=(2\pi)^{-1/2}\int_{-\infty}^\infty dq e^{-ikq}\psi(q)$. This then turns 
	\begin{equation}
		R_c^* = k_o \int_0^\infty d\zeta \mathcal{T}_B(-V_0,\zeta) \int_{-\infty}^\infty dk e^{i k \zeta} \abs{\tilde{\psi}(k)}^2
	\end{equation}
	which is evaluated using the methods of Ref. \cite{PhysRevA.101.022103}. The entire procedure is outlined in the Supplementary Material. Taking the quantity $\Im[R_c^*]= ( \hbar k_o / \mu c) \tilde{R}_c $ yields
	\begin{equation}
		\tilde{R}_c= \tilde{R}_c^{(+)} - \tilde{R}_c^{(-)}
		\label{eq:Rcfinal}
	\end{equation}
	\begin{equation}
		\tilde{R}^{(\pm)}_{c} = \int_{\kappa_c}^\infty dk \abs{\tilde{\psi}(\pm k)}^2 \sqrt{\dfrac{\tilde{E}_k^2}{(\tilde{E}_k-V_o)^2-\mu^2 c^4}}
		\label{eq:RcfinalPM}
	\end{equation}
	wherein,
	\begin{align}
		\kappa_c =& \sqrt{\dfrac{2\mu V_o}{\hbar^2}\left(\frac{V_o}{2\mu c^2}+ 1\right)}\\
		\tilde{E}_k =&\sqrt{\hbar^2k^2c^2 + \mu^2 c^4}.
	\end{align}
	The quantity $\bar{\tau}_{\text{trav}} = t_c \tilde{R}_c$ is then identified as the traversal time across the barrier region, where $t_c=L/c$ is the time it takes a photon to traverse the barrier length. The equivalence of Eqs. \eqref{eq:RCRaw} and \eqref{eq:Rcfinal} has been verified numerically using a Gaussian wavepacket $\varphi(q)=(2\pi\sigma^2)^{-1/4}\exp(-(q-q_o)^2/4\sigma^2)$, and will be detailed elsewhere. In the non-relativistic limit $\bar{\tau}_{\text{trav}}$ reduces to the barrier traversal time derived in Ref. \cite{PhysRevLett.108.170402} using a non-relativistic TOA-operator, i.e. 
	\begin{equation}
		\lim_{c\rightarrow\infty} t_c \tilde{R}_c = \dfrac{\mu  L}{p_o} k_o \int_{\kappa}^\infty dk \dfrac{\abs{\tilde{\psi}(+ k)}^2-\abs{\tilde{\psi}(- k)}^2}{\sqrt{k^2-\kappa^2}}
	\end{equation}
	where, $\kappa=\sqrt{2\mu V_o/\hbar^2}$.
	
	The terms $\tilde{R}_c^{(+)}$ and $\tilde{R}_c^{(-)}$ characterize the contribution of the positive and negative components of the energy distribution of $\tilde{\psi}(k)$ with $\abs{k}>\kappa_c$ to the effective IOR $\tilde{R}_c$, respectively. Thus, the quantity 
	\begin{equation}
		\bar{\tau}_{\text{trav}}^{(\pm)} = t_c\tilde{R}_c^{(\pm)} = \int_{\kappa_c}^\infty dk \bar{\tau}_{\text{top}}(k) |\tilde{\psi}(\pm k)|^2
	\end{equation}
	is the weighted average of the classical above barrier traversal time $\bar{\tau}_{\text{top}}(k) = t_c\tilde{E}_k ( (\tilde{E}_k-V_o)^2-\mu^2 c^4 )^{-1/2}$ with weights $|\tilde{\psi}(\pm k)|^2$. Moreover, Eq. \eqref{eq:RcfinalPM} shows the contribution of the below barrier energy components of $\tilde{\psi}(k)$ with $\abs{k}<\kappa_c$ vanishes, which leads us to the same conclusion as that of Ref. \cite{PhysRevLett.108.170402}. That is, the below barrier energy components of $\tilde{\psi}(k)$ are transmitted instantaneously which implies that tunneling, whenever it occurs, is instantaneous. The same results can be obtained by imposing Born-Jordan ($\alpha_k^{(n)}=1$) and simple symmetric ordering ($\alpha_k^{(n)}=\delta_{k,0}+\delta_{n,k}$) on the quantized relativistic TOA-operators.
	
	In conclusion, the instantaneous tunneling time predicted by Ref. \cite{PhysRevLett.108.170402} is not a mere consequence of using a non-relativistic theory but is an inherent quantum effect in the context of ``arrival times''. However, this instantaneous tunneling time may only be observed for a certain configuration of the experiment. Specifically, it  follows from  Eq. \eqref{eq:RcfinalPM} that the initial incident wavepacket $\psi(q)$ must be sufficiently spatially wide. This will ensure that the spread in momentum is narrow so that $\tilde{\psi}(k)$ only has below barrier components. Moroever, Eq. \eqref{eq:RcfinalPM} rests on the assumption that $\psi(q)$ does not initially `leak' inside the barrier region, as such, the initial incident wavepacket must be placed very far from the barrier.

	P.C.M. Flores acknowledges the support of the Department of Science and Technology - Science Education Institute through the ASTHRDP-NSC graduate scholarship progam. This work was also supported by UP-System Enhanced Creative Work and Research Grant ECWRG 2019-05-R.

	
	\bibliography{reltunnel.bib}
	
	\foreach \x in {1,2,3}
	{%
		\clearpage
		

\section*{Supplementary material}

# Supplementary Material:
# Vanishing of the below barrier components

Philip Caesar M. Flores    Eric A. Galapon

November 22, 2022

For completeness, we provide an outline for the evaluation of the effective index of refraction (IOR) and show the vanishing of the below-barrier contributions to the IOR. Substituting Eq. (17) into Eq. (30) of the main text, the complex-valued IOR now has the form

$$\frac{R_c^*}{k_o} = \int_0^\infty d\zeta \mathcal{T}_B(-V_0,\zeta) \int_{-\infty}^\infty dk e^{ik\zeta} \left|\tilde{\psi}(k)\right|^2 \tag{1}$$

$$= \int_0^\infty d\zeta \mathcal{F}_B(-V_0,\zeta) \int_{-\infty}^\infty dk e^{ik\zeta} \left|\tilde{\psi}(k)\right|^2 + \frac{2}{\pi}\int_1^\infty dy \frac{\sqrt{y^2-1}}{y}\mathcal{G}_B(V_o,z) \int_{-\infty}^\infty dk \frac{\frac{\mu_o c}{\hbar}y}{k^2+\frac{\mu_o^2 c^2}{\hbar^2}y^2}\left|\tilde{\psi}(k)\right|^2. \tag{2}$$

where,

$$\mathcal{F}_B(V_o,\zeta) = \int_0^\infty dy e^{-y} \oint_R \frac{dz}{2\pi i}\frac{1}{z}\sqrt{1+\frac{z^2}{\mu^2 c^2}}{}_0F_1\left[;1;\frac{\mu V_o}{2\hbar^2}\left(\zeta-i\hbar\frac{y}{z}\right)^2 \mathcal{P}_B(V_o,z)\right] \tag{3}$$

$$\mathcal{G}_B(V_o,y) = \frac{1}{2}\left(1-\left(\frac{1}{y}\frac{V_o}{\mu c^2}\right)^2 + 2i\frac{\sqrt{y^2-1}}{y^2}\left(\frac{V_o}{\mu c^2}\right)\right)^{-1/2} + g_{i\to -i}. \tag{4}$$

Eq. (2) follows from interchanging the order of integration in the second term, but the same cannot be done on the first term. Specifically, if we use the series representation of the hypergeometric function in $\mathcal{F}_B(V_o,\zeta)$ to perform a term-by-term integration, then this will lead to an infinite sum of divergent integrals. A finite value for these resulting divergent integrals may be assigned using analytic continuation, regularization, and many others. However, it was recently shown by one of us that this naive interchange in the ordering of integrals leading to divergent integrals sometimes miss significant terms [1, 2]. This was shown to have physical significance in the traversal time of a non-relativistic particle across a potential well [3].

To make the the interchange in the orders of integration on the the first term of Eq. (2) valid, we use the methods of Pablico and Galapon [3] and use the contour shown in Fig. 1. We let $p(z)=|\tilde{\psi}(z)|^2$ and assume that $\tilde{\psi}(z)$ does not have any poles in the complex plane, i.e.,

$$\int_{-\infty}^\infty e^{ix\zeta}p(x) = \int_{-\infty}^\infty dx e^{-(\epsilon-ix)\zeta}p(x+i\epsilon). \tag{5}$$

This now makes

$$\int_0^\infty d\zeta \mathcal{F}_B(-V_0,\zeta)\int_{-\infty}^\infty dk e^{ik\zeta}\left|\tilde{\psi}(k)\right|^2 = \int_{-\infty}^\infty dk p(k+i\epsilon) \int_0^\infty d\zeta \mathcal{F}_B(-V_0,\zeta) e^{-(\epsilon-ik)\zeta}. \tag{6}$$

The interchange is valid provided that $\epsilon > k$. We can now use the series representation of the hypergeometric function in $\mathcal{F}_B(-V_o,\zeta)$ to perform a term-by-term integration. This yields the expansion

$$\frac{R_c^*}{k_o} = i\sum_{n=0}^\infty \frac{(2n)!}{(1)_n n!}\left(\frac{\mu_o V_o}{2\hbar^2}\right)^n \int_{-\infty}^\infty dk p(k+i\epsilon)\left\{\left(\sqrt{1+\frac{\hbar^2(k+i\epsilon)^2}{\mu_o^2 c^2}}\right)^{n+1}\right.$$
$$\left.\times\left((k+i\epsilon)^2+\frac{V_o^2}{\hbar^2 c^2}\right)^{-n-\frac{1}{2}}\mathrm{csgn}(k+i\epsilon)\right\} - 2\int_1^\infty dy\frac{\sqrt{y^2-1}}{y}\mathcal{G}_B(V_o,z)p\left(i\frac{\mu_o c}{\hbar}y\right)$$
$$+\frac{2}{\pi}\int_1^\infty dy\frac{\sqrt{y^2-1}}{y}\mathcal{G}_B(V_o,z)\int_{-\infty}^\infty dk\frac{\frac{\mu_o c}{\hbar}y}{k^2+\frac{\mu_o^2 c^2}{\hbar^2}y^2}p(k) \tag{7}$$

1

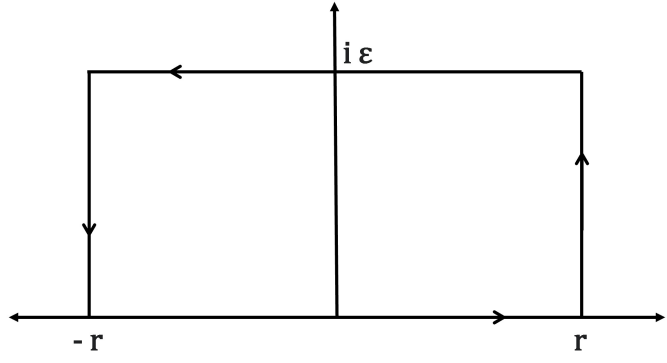

Figure 1: Contour of integration for Eq. (5) leading to the interchange of the order of integration in Eq. (6)

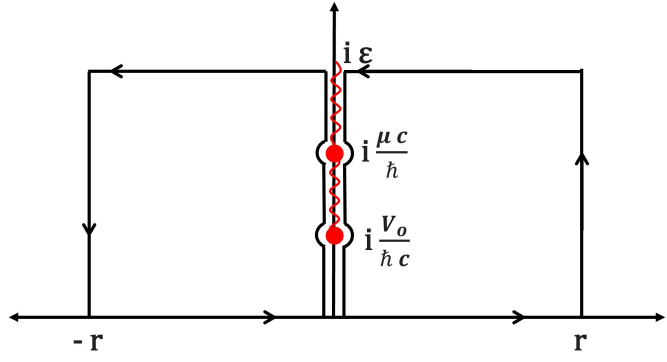

Figure 2: Contour of integration for the series expansion in Eq. (7), when $V_o/\hbar c < \mu_o c/\hbar$.

To understand the physical content of Eq. (7), we consider the following integral in the complex plane,

$$\oint dz p(z) \left(\sqrt{1 + \frac{\hbar^2 z^2}{\mu_o^2 c^2}}\right)^{n+1} \left(z^2 + \frac{V_o^2}{\hbar^2 c^2}\right)^{-n-\frac{1}{2}}, \tag{8}$$

which has four branch points at $z = \{\pm i\frac{\mu_o c}{\hbar}, \pm i\frac{V_o}{\hbar c}\}$. We assume that the branch points satisfy $V_o/\hbar c < \mu_o c/\hbar$ which is equivalent to the condition $V_o < \mu_o c^2$, and evaluate Eq. (8) using the contours in Fig. 2. Thus, we have

$$\int_{-\infty}^{\infty} dk p(k+i\epsilon) \left(\sqrt{1 + \frac{\hbar^2(k+i\epsilon)^2}{\mu_o^2 c^2}}\right)^{n+1} \left((k+i\epsilon)^2 + \frac{V_o^2}{\hbar^2 c^2}\right)^{-n-\frac{1}{2}} \operatorname{csgn}(k+i\epsilon)$$

$$= \int_0^{\infty} dk (p(k) - p(-k)) \left(\sqrt{1 + \frac{\hbar^2 k^2}{\mu_o^2 c^2}}\right)^{n+1} \left(x^2 + \frac{V_o^2}{\hbar^2 c^2}\right)^{-n-\frac{1}{2}}$$

$$+ (-i)^{n+1} \left(\frac{\hbar^2}{\mu_o^2 c^2}\right)^n (1 + (-1)^n) \int_1^{\infty} dy \frac{\left(\sqrt{y^2 - 1}\right)^{n+1}}{\left(y^2 - \frac{V_o^2}{\mu_o^2 c^4}\right)^{n+\frac{1}{2}}} p\left(i\frac{\mu_o c}{\hbar} y\right). \tag{9}$$

where $\operatorname{csgn}(z)$ is the complex signum function

$$\operatorname{csgn}(z) = \begin{cases} 1 & , \operatorname{Re}(z) > 0 \\ -1 & , \operatorname{Re}(z) < 0 \\ \operatorname{sgn}(\operatorname{Im}(z)) & , \operatorname{Re}(z) = 0. \end{cases} \tag{10}$$

2

We then substitute Eq. (9) to Eq. (7), and interchange the orders of summation and integration. Taking the imaginary component of the IOR yeilds

$$\text{Im}[R_c^*] = \frac{\hbar k_o}{\mu_o c}\tilde{R}_c = \frac{\hbar k_o}{\mu_o c}\text{Re}\left\{\int_0^\infty dk\left(\left|\tilde{\psi}(k)\right|^2 - \left|\tilde{\psi}(-k)\right|^2\right)\sqrt{\frac{\tilde{E}_k^2}{(\tilde{E}_k - V_o)^2 - \mu_o^2 c^4}}\right\} \quad (11)$$

in which, $\tilde{E}_k = \sqrt{\hbar^2 k^2 c^2 + \mu_o^2 c^4}$. The right-hand side of Eq. (11) is only real-valued when $|k| > \kappa_c$, where

$$\kappa_c = \sqrt{\frac{2\mu_o V_o}{\hbar^2}\left(1 + \frac{V_o}{2\mu_o c^2}\right)} \quad (12)$$

provided that $V_o < \mu_o c^2$. Thus, the IOR now has the form

$$\tilde{R}_c = \tilde{R}_c^{(+)} - \tilde{R}_c^{(-)} = \int_{\kappa_c}^\infty dk\left|\tilde{\psi}(+k)\right|^2\sqrt{\frac{\tilde{E}_k^2}{(\tilde{E}_k - V_o)^2 - \mu_o^2 c^4}} - \int_{\kappa_c}^\infty dk\left|\tilde{\psi}(-k)\right|^2\sqrt{\frac{\tilde{E}_k^2}{(\tilde{E}_k - V_o)^2 - \mu_o^2 c^4}} \quad (13)$$


	}
	
\end{document}